\documentclass[12pt]{article}
\usepackage{amsfonts}
\usepackage{bbm}
\usepackage{graphicx}
\usepackage{pgfplots}
\usepackage{booktabs}
\usepackage{subcaption}
\usepackage{mathrsfs}
\usepackage{color}
\usepackage{verbatim}
\usepackage{cancel}
\usepackage{mhchem} 
\usepackage{geometry}             
\usepackage{amssymb}
\usepackage{amsmath}
\usepackage{epstopdf}
\usepackage{cases}
\usepackage{cite}
\usepackage{setspace}

\usepackage[hyperindex=true,
pdfstartview=FitH,
bookmarksnumbered=true,
bookmarksopen=true,
citecolor=blue,
linkcolor=blue,
colorlinks=true,
urlcolor=rossoCP3,
unicode]{hyperref}

\definecolor{rossoCP3}{cmyk}{0,.88,.77,.40}

\renewcommand{\Re}{\operatorname{Re}} 
\renewcommand{\Im}{\operatorname{Im}} 

\begin{document}
	
	\title{\bf
		Thermodynamic supercriticality and complex phase diagram for charged Gauss-Bonnet AdS black holes}
	{\author{\small  Zhi-Yuan Li${}^{1}$, Xuan-Rui Chen${}^{1}$, Bin Wu${}^{1,2,3}$\thanks{{\em email}: \href{mailto:binwu@nwu.edu.cn}{binwu@nwu.edu.cn}}{ }, Zhen-Ming Xu${}^{1,2,3}$
			\vspace{5pt}\\
			\small $^{1}${\it School of Physics, Northwest University, Xi'an 710127, China}\\
			\small $^{2}${\it Shaanxi Key Laboratory for Theoretical Physics Frontiers, Xi'an 710127, China}\\
			\small $^{3}${\it Peng Huanwu Center for Fundamental Theory, Xi'an 710127, China}}
		
		\date{}
		\maketitle
		\begin{spacing}{1.15}
			\begin{abstract}
				Lee-Yang zero theory plays a crucial role in phase transition theory and is widely employed in the critical behavior of statistical thermodynamics. The supercritical regime of black hole thermodynamics remains a relatively unexplored area, and recent applications of this theory to charged anti-de Sitter (AdS) black holes have initiated probes into this regime, revealing a simple structure partitioned by a single Widom line. In this paper, we apply Lee-Yang theory to charged Gauss-Bonnet AdS black holes, which feature complex phase diagrams (e.g., triple points), to determine how such structures are reflected in the supercritical regime. Notably, we observe a key dimensional difference in that the five-dimensional (5d) system, which lacks a triple point, confirms the known single Widom line structure, while the six-dimensional (6d) system, which admits a triple point, generates two distinct Widom lines. These lines partition the supercritical domain into three sectors (small-, intermediate-, and large- black hole-like phases), corresponding to the three phases coexisting at the triple point. Our results reveal a direct  correspondence between the number of coexisting phases at a triple point and the number of distinct supercritical sectors separated by Widom lines.
			\end{abstract}
			
			\section{Introduction}	
			
			Phase transition is one of the core concepts in thermodynamics. In classical thermodynamics, the mutual transformation of substances between different phases has been widely studied since the 19th century. It reveals the fundamental laws governing the collective behavior of physical systems when external conditions change. Phase transition processes are usually accompanied by sudden changes in certain physical quantities and the emergence of critical behaviors, which form the basis for understanding the behavior of many complex systems.

		    Due to the universality of thermodynamics, applying thermodynamic methods to gravitational systems, particularly black hole systems, has become a natural exploration direction. Studies have shown that black holes, similar to classical thermodynamic systems, also exhibit phase transition behaviors. The first discovered phase transition was the Hawking–Page transition\cite{Hawking:1982dh}. By analyzing the zero point of the Gibbs free energy of black holes, Hawking and Page revealed the phase transition process from the thermal radiation state in the AdS space to the black hole state. Subsequent studies have shown that the phase transition behavior of charged AdS black holes has significant similarities with classical thermodynamic systems such as Van der Waals (VdW) fluids\cite{Chamblin:1999tk,Chamblin:1999hg}. Furthermore, in the extended phase space thermodynamics developed by interpreting the negative cosmological constant $\Lambda$ as pressure $P$ and the conjugate quantity as the thermal volume $V$, it has been discovered that black holes exhibit more complex thermodynamic behaviors\cite{Kastor:2009wy,Dolan:2010ha,Cvetic:2010jb,Dolan:2011xt,Kubiznak:2012wp}, and this framework is also known as black hole chemistry\cite{Frassino:2015oca,Kubiznak:2016qmn}. We refer to the article \cite{Mann:2025xrb} and the references cited therein for a recent comprehensive review.

	    	For a long time, the study of black hole thermodynamics has mainly focused on phenomena below the critical point. In this region, universal critical behaviors can be observed, and phase transitions usually end within this range\cite{Kubiznak:2012wp, Kubiznak:2016qmn}. To fully understand the thermodynamic properties of black holes, attention has gradually shifted to the supercritical regime. Classical thermodynamics has achieved many results in studying the supercritical regions, such as the research on fluids like water and \ce{CO2}\cite{simeoni2010widom, 10.1063/1.4930542,NISHIKAWA1995149,doi:10.1021/acs.jpcb.9b04058, li2024thermodynamic}. For systems with only one critical point, such as VdW fluids, the supercritical region is usually divided into liquid-like and gas-like regions by a Widom line, which extends outward from the critical point and is defined as the locus of extrema of response functions such as heat capacity, representing a smooth continuation of phase transition behavior\cite{G.O.Jones_1956,xu2005relation,PhysRevE.86.052103,PhysRevLett.112.135701,PhysRevE.95.052120}. In contrast, systems with multiple critical points or triple points often exhibit more complex supercritical structures, possibly with several Widom lines and different coexistence regions\cite{li2024thermodynamic}.

            In black hole thermodynamics, the supercritical regime has also been gaining more interest. The Lee–Yang phase transition theory provides a new perspective for studying the critical properties of gravitational systems\cite{Yang:1952be,Lee:1952ig}. 
            This theory originates from statistical physics, and its core idea is to reveal the essence of phase transitions by studying the distribution law of the zeros of the partition function (which correspond to singularities of the Gibbs free energy) on the complex plane. Early research proposed preliminary concepts regarding the thermodynamic geometry of AdS black holes above the critical point\cite{Sahay:2017hlq}. Further, in the literatures\cite{Zhao:2025ecg,Xu:2025jrk,Wang:2025ctk}, the authors explored the supercritical characteristics in various black hole systems. Recently, the Lee–Yang phase transition theory has been applied to construct the complex phase diagram of four-dimensional charged AdS black holes and identified a Widom line, which divides the supercritical region into small- and large- black hole–like phases (SBH-like/LBH-like) \cite{Xu:2025jrk}.

	        Despite these advances, it remains unclear whether black hole systems with more complex phase structures exhibit distinct phase regions in the supercritical regime. Gauss–Bonnet gravity, as a higher-dimensional generalization of Einstein gravity\cite{Lovelock:1971yv,Boulware:1985wk,Cai:2001dz}, provides an ideal model to investigate this question. In a 5-dimensional Gauss-Bonnet AdS black hole system, there is usually a critical point, and its thermodynamic behavior is similar to that of the classical VdW fluid\cite{Cvetic:2001bk,Cai:2013qga,Xu:2013zea}. In contrast, a 6-dimensional system with a specific coupling value exhibits a triple point, allowing small, intermediate, and large black holes (SBH/IBH/LBH) to coexist simultaneously\cite{Wei:2014hba, Frassino:2014pha}. In this paper, we will utilize the Lee–Yang phase transition theory to conduct a systematic investigation of the supercritical behavior of charged Gauss-Bonnet AdS black holes in both 5d and 6d. Our objective is to determine how the increased complexity of the phase diagram, especially at the triple point in 6d, manifests in the structure of the supercritical region. We hope to provide a more comprehensive understanding of the underlying mechanisms of phase transitions in gravitational systems.

	        The organization of this paper is as follows: In Section \ref{II}, we review the thermodynamics of charged Gauss-Bonnet AdS black holes. Furthermore, we analyze the selection of parameters and present a scheme for reducing thermodynamic quantities. Then, in Sections \ref{III} and \ref{IV}, the complex singularities of Gibbs free energy in the 5d and 6d cases were calculated. We constructed complex phase diagrams in two cases and analyzed their supercritical characteristics. Finally, our conclusion is presented in Section \ref{V}.

			\section{Thermodynamics of charged Gauss-Bonnet AdS black holes}\label{II}
			
			The action of the Einstein-Gauss-Bonnet gravity in $d$-dimensional spacetime can be written as\cite{Cvetic:2001bk,Cai:2001dz,Cai:2013qga}
			\begin{equation}
				S=\frac{1}{16\pi G_{d}} \int d^{d}x\sqrt{-g}
				\Big[\mathcal{R}
				-2\Lambda
				+\alpha_{\text{GB}}\mathcal{L}_{\text{GB}}
				-\mu_0 \mathcal{F}_{\mu \nu} \mathcal{F}^{\mu \nu}\Big], \footnote{The constant $G_d$ and $\mu_0$ are retained here, and their value are taken to be consistent with the convention adopted in Ref.\cite{Cvetic:2001bk,Cai:2001dz,Cai:2013qga}.}
				\label{action}
			\end{equation}
			where $\Lambda$ is the cosmological constant, and $\alpha_{\text{GB}}$ is the Gauss-Bonnet coupling constant. The Gauss-Bonnet Lagrangian $\mathcal{L}_{\text{GB}}$ and the electromagnetic Lagrangian $\mathcal{L}_\text {matter }$ take the form
			\begin{equation}
				\mathcal{L}_{\mathrm{GB}}=\mathcal{R}_{\mu \nu \gamma \delta} \mathcal{R}^{\mu \nu \gamma \delta}-4 \mathcal{R}_{\mu \nu} \mathcal{R}^{\mu \nu}+\mathcal{R}^2, \qquad
				\end{equation}
			with the Maxwell field strength defined as
			$\mathcal{F}_{\mu\nu}=\partial_{\mu}\mathcal{A}_{\nu}-\partial_{\nu}\mathcal{A}_{\mu}$.
			
			The static, spherically symmetric black hole solution for the action (\ref{action}) is
			\begin{equation}
			ds^{2}=-f(r)dt^{2}+f^{-1}(r)dr^{2}+r^{2}d\Omega_{d-2}^{2},
			\end{equation}
			where the metric function $f(r)$ is given by\cite{Wiltshire:1985us,Kofinas:2006hr,Hendi:2015pda}
			\begin{equation}
			f(r)=1+\frac{r^2}{2\alpha}\left[1-\sqrt{1+\frac{2\alpha}{d-2}\left(\frac{32\pi M}{\Sigma_{d-2}r^{d-1}}-\frac{4Q^2}{(d-3)r^{2d-4}}+\frac{4\Lambda}{d-1}\right)}\right].
			\end{equation}
			Here, $M$ and $Q$ are the black hole mass and charge, respectively. The parameter $\alpha=(d-3)(d-4)\alpha_{\text{GB}}$ is the effective Gauss-Bonnet coupling constant, and $\Sigma_{d-2}=2\pi^{\frac{d-1}{2}}/\Gamma[(d-1)/2]$ is the area of a unit $(d-2)$-dimensional sphere.
			
			In the extended phase space, the thermodynamic pressure of an AdS black hole is associated with the negative cosmological constant  as\cite{Kastor:2009wy,Kubiznak:2016qmn}
			\begin{equation}
			P=-\frac{1}{8\pi }\Lambda.
			\end{equation}
			To ensure a well-defined vacuum solution with $M=Q=0$, the pressure $P$ and the effective Gauss-Bonnet coupling constant $\alpha$ must satisfy the bound\cite{Cai:2013qga}
			\begin{equation}\label{prg}
			0 <\frac{64\pi\alpha P}{(d-1)(d-2)}\leq 1,\quad d\geq5.
			\end{equation}
			
			The event horizon radius $r_{h}$ is the largest real root of $f(r_{h})=0$. The mass M (interpreted as enthalpy $M \equiv H$ in the extended phase space thermodynamics), temperature $T$, and entropy $S$ are given by\cite{Cai:2013qga,Xu:2013zea,Wei:2014hba}
			\begin{align}
				M &= \frac{Q^2\Sigma_{d-2}}{8\pi(d-3)r_h^{d-3}}+\frac{Pr_h^{d-1}\Sigma_{d-2}}{d-1}+\frac{(d-2)r_h^{d-3}\Sigma_{d-2}}{16\pi}\left(1+\frac{\alpha}{r_h^2}\right), \label{eq:M} \\
				T &= \frac{-2Q^{2}r_{h}^{8-2d}+(d-3)(d-2)r_{h}^{2}+(d-5)(d-2)\alpha+16\pi Pr_{h}^{4}}{4\pi(d-2)r_{h}(r_{h}^{2}+2\alpha)}, \label{eq:T} \\
				S &= \frac{r_h^{d-2}\Sigma_{d-2}}{4}\left(1+\frac{2(d-2)\alpha}{(d-4)r_h^2}\right). \label{eq:S}
			\end{align}
			The Gibbs free energy $G = M - TS$ is given by the expression
			\begin{equation} \label{eq:G}
				\begin{split}
					G =& \frac{\Sigma_{d-2}}{16\pi}r_h^{-d-5}\Bigg[r_h^{2d}\left(\alpha(d-2)+\frac{16\pi Pr_h^4}{d-1}+(d-2)r_h^2\right) + \frac{2Q^2r_h^8}{d-3} \\
					& -\frac{r_h^2\left(\frac{2\alpha(d-2)}{(d-4)r_h^2}+1\right)\left(r_h^{2d}\left(\alpha(d-5)(d-2)+(d-3)(d-2)r_h^2+16\pi Pr_h^4\right)-2Q^2r_h^8\right)}{(d-2)\left(2\alpha+r_h^2\right)}\Bigg].
				\end{split}
			\end{equation}
			Finally, the heat capacity $C_P$ is
		    \begin{equation}
					C_P = T\left(\frac{\partial S}{\partial T}\right)_P = T{{\left( \frac{\partial S}{\partial {{r}_{h}}} \right)}_{P}}{{\left( \frac{\partial {{r}_{h}}}{\partial T} \right)}_{P}},
			\end{equation}
			where
			\begin{eqnarray}
				\frac{\partial S}{\partial r_h} &=&
				\frac{(d-2){{\Sigma }_{d-2}}}{4}{{r}_{h}}^{d-5}\left( {{r}_{h}}^{2}+2\alpha  \right), \nonumber \\
				\frac{\partial T}{\partial r_h}
				&=& \frac{1}{4\pi (d-2){r^2}{{\left( {{r}^{2}}+2\alpha  \right)}^{2}}}
				\left [ 16\pi P{{r}^{6}}-{{r}^{4}}((d-5)d-96\pi \alpha P+6)-\alpha (d-9)(d-2){{r}^{2}}
				\right. \nonumber \\
				&&+2{{Q}^{2}}{{r}^{8-2d}}\left( 2\alpha (2d-7)+(2d-5){{r}^{2}} \right)-2{{\alpha }^{2}}(d-5)(d-2) ].
			\end{eqnarray}

			To simplify the analysis and facilitate a universal discussion, a convenient and standard approach is to rescale the thermodynamic quantities by the black hole charge $Q$\cite{Cai:2013qga}. The thermodynamic quantities scale with $Q$ as
			\begin{equation}
				\label{eq:scaling} 
				r_{h} \propto |Q|^{\frac{1}{d-3}},\quad
				\alpha \propto |Q|^{\frac{2}{d-3}},\quad
				P \propto |Q|^{\frac{-2}{d-3}},\quad
				T \propto |Q|^{\frac{-1}{d-3}},\quad
				G \propto |Q|.
			\end{equation}
			
			Since this scaling property eliminates the dependence on the charge $Q$, we can set $Q=1$ without loss of generality in our subsequent analysis, other nonzero values of $Q$ only rescale the thermodynamic variables but do not change the underlying phase structure\cite{Wei:2014hba}. With $Q$ fixed, the thermodynamic phase behavior is governed by the effective Gauss-Bonnet coupling constant $\alpha$. The influence of $\alpha$ differs significantly between dimensions, providing the central contrast for our investigation. 
			
			As analyzed in Ref.\cite{Wei_2022}, the 6d case can exhibit a complex phase structure (specifically a triple point where SBH/IBH/LBH coexist) when $\alpha$ falls within a specific parameter range. This triple point is enabled by a non-monotonic $T(r_{h})$ isobar, which develops two local maxima and two local minima (corresponding to four real roots of $(\partial_{r_{h}} T)_{P} = 0$). This structure permits two distinct first-order phase transitions (SBH-IBH and IBH-LBH) to be mediated by the Maxwell equal area law\cite{SPALLUCCI2013436}. In contrast, the 5d system invariably retains a simpler VdW-like phase structure with only a single critical point\cite{Cvetic:2001bk,Cai:2013qga,Xu:2013zea}. To implement a direct comparison based on this contrast, we follow the precedent of Ref.\cite{Wei_2022} and fix $\alpha = 3.05$.

			We now proceed by reducing the thermodynamic quantities in the 5d and 6d systems, normalizing them relative to the characteristic points of each system. We define the reduced pressure, temperature, horizon radius, Gibbs free energy and heat capacity as $p$, $t$, $z$, $g$ and $c_p$, respectively.
		
			\begin{itemize}
				\item For the 5d case (single critical point), the quantities are reduced relative to the critical point ($P_c, T_c, r_c, G_c$) as
				     \begin{equation}\label{5dr}
					       p = \frac{P}{P_c}, \quad t = \frac{T}{T_c}, \quad z = \frac{r_h}{r_c}, \quad g = \frac{G}{G_c}, \quad c_p = C_P\cdot P_c.
				     \end{equation}
				The critical point parameters are determined by simultaneously solving the conditions $(\partial_{r_{h}} T)_{P} = 0$ and $(\partial_{r_{h}, r_{h}} T)_{P} = 0$. Their numerical values will be provided and discussed in the next Section \ref{III}.
				
				\item For the 6d case (triple point), the quantities are reduced relative to the triple point ($P_t, T_t, r_t, G_t$) as
				     \begin{equation}\label{6dr}
						   p = \frac{P}{P_t}, \quad t = \frac{T}{T_t}, \quad z = \frac{r_h}{r_t}, \quad g = \frac{G}{G_t}, \quad c_p = C_P\cdot P_t.
				     \end{equation}
				The triple point is precisely determined by the thermodynamic requirement that the Gibbs free energy $G$ must be equal for all three phases ($G_{\text{SBH}} = G_{\text{IBH}} = G_{\text{LBH}}$)\cite{Wei:2014hba,Frassino:2014pha}. The numerical values associated with this triple point and its related critical points will be given in the Section \ref{IV}.
				     
			\end{itemize} 
			
			This comprehensive reduced formulation allows for a unified and comparative analysis of the thermodynamic behavior across different dimensions and pressure regimes, while also highlighting the universal features of the system.

			\section{Phase structure and supercriticality in five-dimensions}\label{III}
			In order to gradually understand the supercritical behavior and the formation of complex phase structures, we first analyze the 5d system. This system has a single critical point and its phase transition behavior is relatively simple, similar to a VdW fluid.	
			\subsection{Reduced equations of state and critical behavior}
			
			Using the reduction scheme defined in (\ref{5dr}), with $Q=1, \alpha=3.05$, and reducing relative to the 5d critical point, the resulting critical parameters are\cite{Cai:2013qga,Wei:2014hba}
			\begin{equation}\label{5dc}
				P_c \approx 0.002156, \quad T_c \approx 0.0371, \quad r_c \approx 4.327, \quad G_c \approx 0.0741.
			\end{equation}
			The corresponding reduced thermodynamic quantities can then be expressed as
			\begin{equation}\label{seq1}
				\begin{split}
					&t=\frac{8p{{P}_{c}}\pi {{r}_{c}}^{6}{{z}^{6}}+3{{r}_{c}}^{4}{{z}^{4}}-{{Q}^{2}}}{6\pi {{r}_{c}}^{5}{{T}_{c}}{{z}^{5}}+12\pi {{r}_{c}}^{3}{{T}_{c}}{{z}^{3}}\alpha },  \\ 	
					&g=\frac{1}{24{{G}_{c}}{{r}_{c}}^{2}{{z}^{2}}\left( {{r}_{c}}^{2}{{z}^{2}}+2\alpha \right)} \left[ -4p{{P}_{c}}{{\pi }^{2}}{{r}_{c}}^{8}{{z}^{8}} - \pi {{r}_{c}}^{6}{{z}^{6}}\left( -3+72p{{P}_{c}}\pi \alpha \right) - 9\pi {{r}_{c}}^{4}{{z}^{4}}\alpha \right. \\
					&\left. \quad + 18\pi {{r}_{c}}^{2}{{z}^{2}}{{\alpha }^{2}} + \pi {{Q}^{2}}\left( 5{{r}_{c}}^{2}{{z}^{2}}+18\alpha \right) \right], \\
					&{{c}_{p}}=\frac{3{{\pi }^{2}}{{P}_{c}}{{r}_{c}}z{{\left( 2\alpha +{{r}_{c}}^{2}{{z}^{2}} \right)}^{2}}\left( 8\pi p{{P}_{c}}{{r}_{c}}^{6}{{z}^{6}}-{{Q}^{2}}+3{{r}_{c}}^{4}{{z}^{4}} \right)}{2\left( 8\pi p{{P}_{c}}{{r}_{c}}^{8}{{z}^{8}}+{{r}_{c}}^{6}{{z}^{6}}(48\pi \alpha p{{P}_{c}}-3)+{{Q}^{2}}\left( 6\alpha +5{{r}_{c}}^{2}{{z}^{2}} \right)+6\alpha {{r}_{c}}^{4}{{z}^{4}} \right)}.  \\ 
				\end{split}				
			\end{equation}
			Based on the above equation, the thermodynamic phase diagram can be constructed, as shown in Fig.\ref{fig1}.

			\begin{figure}[htbp]
				\centering
				\begin{subfigure}[b]{0.48\textwidth}
					\includegraphics[width=\textwidth]{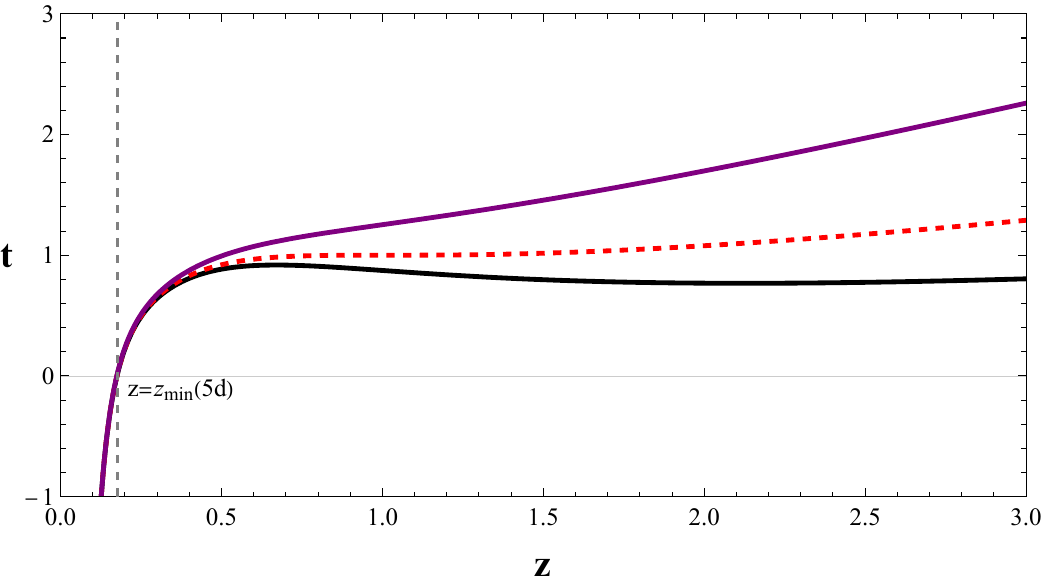}
					\caption{Temperature ($t$) vs. Horizon radius ($z$)}
					\label{fig1a}
				\end{subfigure}
				\hfill
				\begin{subfigure}[b]{0.48\textwidth}
					\includegraphics[width=\textwidth]{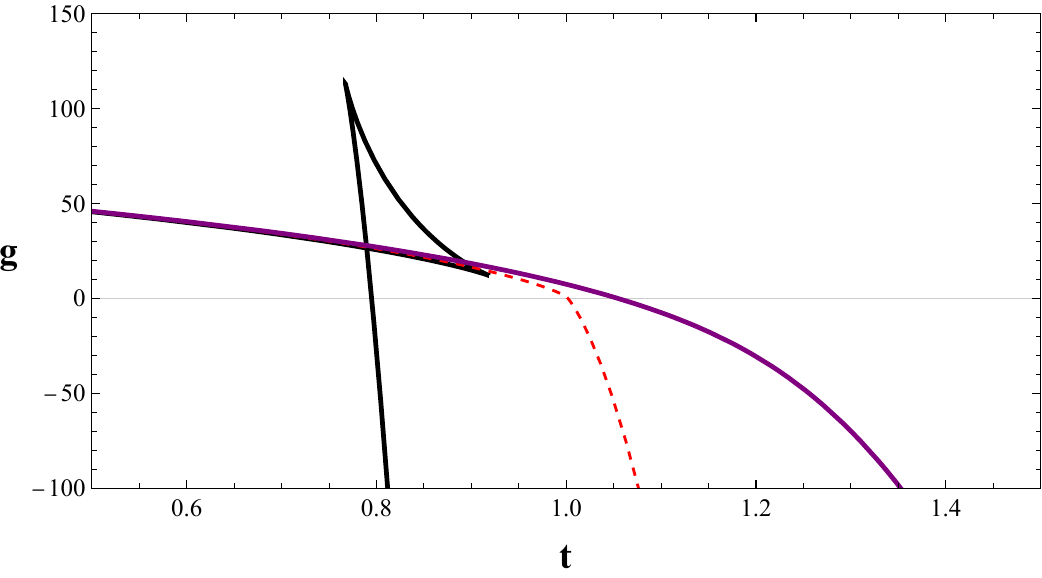}
					\caption{Gibbs free energy ($g$) vs. Temperature ($t$)}
					\label{fig1b}
				\end{subfigure} \\
				
				\begin{subfigure}[b]{0.48\textwidth}
					\includegraphics[width=\textwidth]{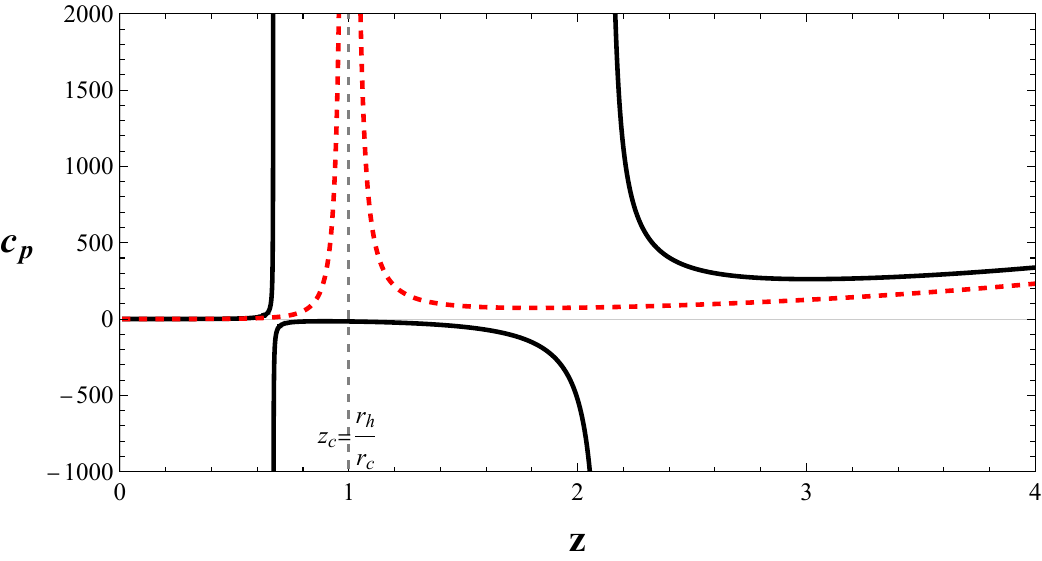}
					\caption{Heat capacity ($c_p$) for $p \le 1$}
					\label{fig1c}
				\end{subfigure}
				\hfill
				\begin{subfigure}[b]{0.48\textwidth}
					\includegraphics[width=\textwidth]{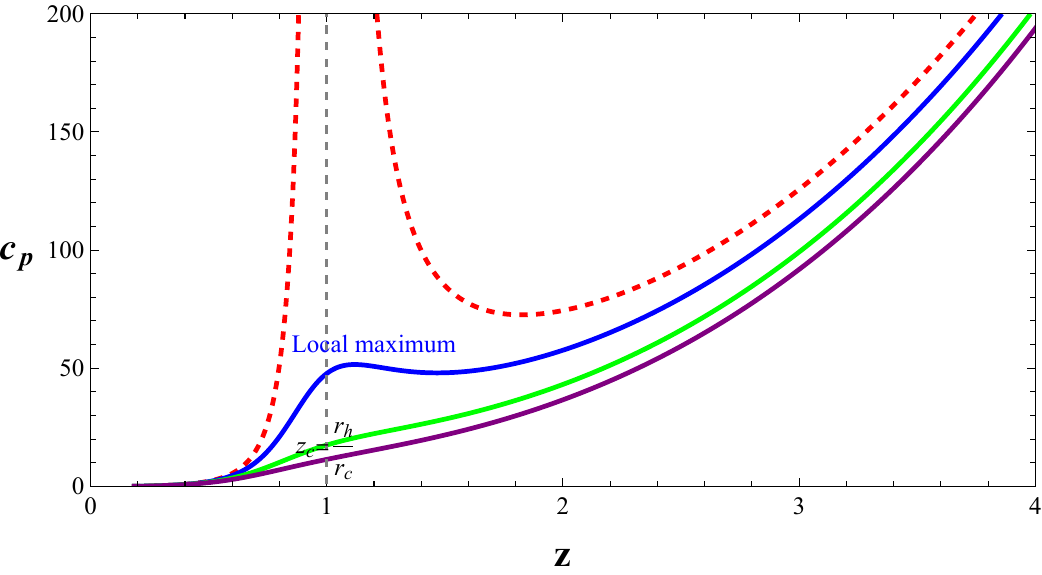}
					\caption{Heat capacity ($c_p$) for $p \ge 1$}
					\label{fig1d}
				\end{subfigure}
				
				\caption{Thermodynamic behavior in 5d charged Gauss-Bonnet AdS black hole ($Q=1, \alpha=3.05$). Curves correspond to different reduced pressures: $p=0.50$ (black), $p=1.00$ (red dashed, critical point), $p=1.20$ (blue), $p=1.60$ (green), $p=2.00$ (purple). The vertical dashed line in (a) indicates the physical boundary $z_{\text{min}}(\text{5d}) \approx 0.176$, below which $t<0$. The critical point ($p=1$) corresponds to the critical radius $z_c=1$.}
				\label{fig1}
			\end{figure}
			
			Fig.~\ref{fig1a} displays the reduced temperature $t$ as a function of the reduced horizon radius $z$. For small values of $z$, the temperature $t$ becomes negative, which is unphysical result. We therefore restrict our analysis to the physical region where $t > 0$. This condition defines a minimum horizon radius $z_{\text{min}}(\text{5d})$, which is indicated by the vertical dashed line in Fig.~\ref{fig1a}. Additionally, the pressure is constrained from above to ensure a well-defined vacuum solution, as given by Eq.(\ref{prg})
			
			\begin{equation}\label{prg1}
				0 < P \le \frac{(d-1)(d-2)}{64\pi \alpha} = \frac{3}{16\pi \alpha} \equiv P_{\text{max}} \quad \Rightarrow \quad 0 < p \le \frac{P_{\text{max}}}{P_c}.
			\end{equation}
			
			With these physical constraints established, we now examine the thermodynamic behavior within this parameter space. For pressures below the critical point $p < 1$, the Gibbs free energy $g$  in Fig.~\ref{fig1b} exhibits a characteristic swallowtail structure, signaling a first-order phase transition between SBH and LBH. This is corroborated by the behavior of the heat capacity $c_p$ in Fig.~\ref{fig1c}. In this range (e.g., the black curve for $p=0.5$), $c_p$ diverges at two distinct horizon radius. At the critical point ($p=1$, red dashed curve), these two divergence points merge into one, occurring precisely at the critical radius $z_c$.
			This well-known VdW-like thermodynamic portrait, characterized by a single critical point, is in full agreement with the established phase behavior described in Refs.\cite{Cvetic:2001bk,Cai:2013qga,Xu:2013zea,Wei:2014hba}. 
			
			As $p$ increases into the critical and supercritical regimes ($p \ge 1$), the swallowtail in the $g-t$ curve vanishes, becoming a smooth, single-valued function. Concurrently, the behavior of $c_p$ changes, the divergence is replaced by a finite local maximum, as shown in Fig.~\ref{fig1d} for $p=1.2, 1.6,$ and $2.0$. 	This signals a smooth crossover, characteristic of the supercritical region. However, this is where the conventional analysis fails for black hole system. In conventional fluid thermodynamics, the Widom line is rigorously defined as the locus of global maxima of a response function, such as the heat capacity $c_p$ \cite{G.O.Jones_1956,xu2005relation,PhysRevE.86.052103,PhysRevLett.112.135701,PhysRevE.95.052120}. This definition is inapplicable to the black hole thermodynamics.
			
			As shown clearly in Fig.~\ref{fig1d}, the supercritical $c_p$ isobars do not possess a global maximum. They only exhibit a finite local maximum, which diminishes in amplitude and broadens significantly as pressure increases. This local feature does not serve as a robust marker for a thermodynamic crossover, rendering the conventional $c_p$-peak definition of the Widom line ambiguous for this system. Therefore, a more fundamental definition is required, one that is not tied to the specific features of $c_p$ peaks. The Lee-Yang theory, as proposed for black holes in Ref.\cite{Xu:2025jrk}, provides exactly such a rigorous alternative by identifying the Widom line with the analytic continuation of the phase transition's singularities in the complex plane. We explore this next.

			\subsection{Complex phase diagram and the Widom line}
			
			To analyze the supercritical regime, we apply the Lee-Yang phase transition theory and extend our analysis to the complex domain\cite{Yang:1952be,Lee:1952ig}. The core idea of this theory is that the thermodynamic properties of a system are encoded in the distribution of the zeros of its partition function $Z$ (the Lee-Yang zeros) in the complex plane. In the thermodynamic limit, the non-analytic behavior characteristic of a phase transition arises as these complex zeros converge onto the real axis. This formalism is well-grounded in black hole thermodynamics. In the semi-classical (Euclidean) framework, the partition function $Z$ is dominated by the Euclidean gravitational action $I_E$, such that $Z \approx e^{-I_E}$. Concurrently, the Gibbs free energy $G$ is identified as $G = T \times I_E$. This directly yields the standard thermodynamic relation $G \approx -T \ln Z$. Therefore, the Lee-Yang zeros of the partition function ($Z=0$) correspond precisely to the singularities of the Gibbs free energy $G$ (where $\ln Z$ diverges).
			
			In our thermodynamic system, a singularity in $G$ manifests physically as a divergence in its second derivatives, most notably the heat capacity $c_p \propto (\partial^2 g / \partial t^2)_p$. The expression for $c_p$ (e.g., Eq.~(\ref{seq1})) is a function of the reduced horizon radius $z$. Consequently, seeking the Lee-Yang singularities is equivalent to solving for the roots $z$ of the $c_p$ divergence condition (i.e., where its denominator is zero). This naturally leads us to extend the analysis to the complex $z$-plane, as these roots are not always real.  They are real for pressures at or below the critical point ($p \le 1$), corresponding to the spinodal lines, but they move into the complex plane in the supercritical region.
			
			Crucially, we do not posit that the horizon radius $z$ becomes physically complex. Rather, the analytic continuation of thermodynamic functions such as $g(z)$ into the complex $z$-plane is employed as a purely mathematical method. The utility of the Lee-Yang theory lies precisely in the fact that the phase transitions on the real axis (such as the SBH-LBH phase transition) are understood as these complex singularities squeezing the real axis within the thermodynamic limit. In the supercritical region, the absence of a first-order phase transition corresponds to these singularities moving off the real axis. The Widom line, representing a smooth crossover, is then physically interpreted as the projection of the trajectories of these complex singularities (specifically, those closest to the real axis) onto the real $(\Re(p), \Re(t))$ plane. This specific application of Lee-Yang theory to rigorously define the Widom line by analyzing complex singularities in charged AdS black holes was recently utilized in Ref.\cite{Xu:2025jrk}.

			\begin{figure}[htbp]
				\centering
				\includegraphics[width=110 mm]{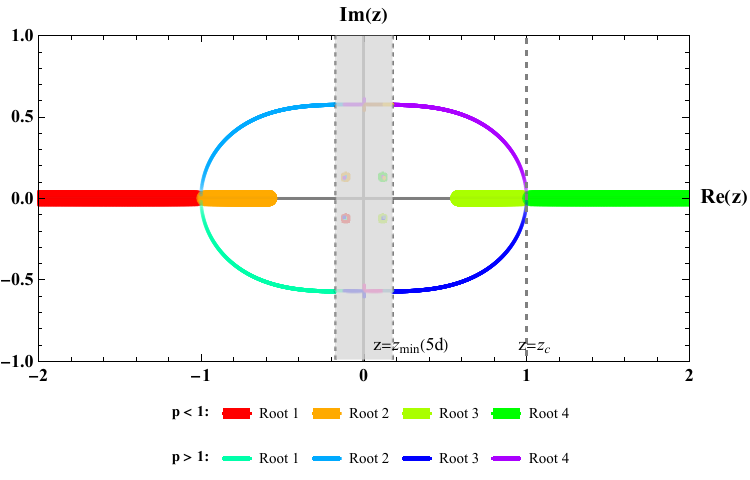}
				\caption{The singularity distribution for the 5d charged Gauss-Bonnet AdS black hole. The distinct colors denote the four physical roots of Eq.~(\ref{fenbu1}). The gray shaded region ($|z| < z_{\text{min}}(\text{5d})$) is non-physical, and roots within it are excluded. The vertical dashed line marks the critical point $z_c$ on the real axis.	
				}
				\label{fig2}	
			\end{figure}
			
			\begin{figure}[htbp]
				\centering
				\includegraphics[width=150 mm]{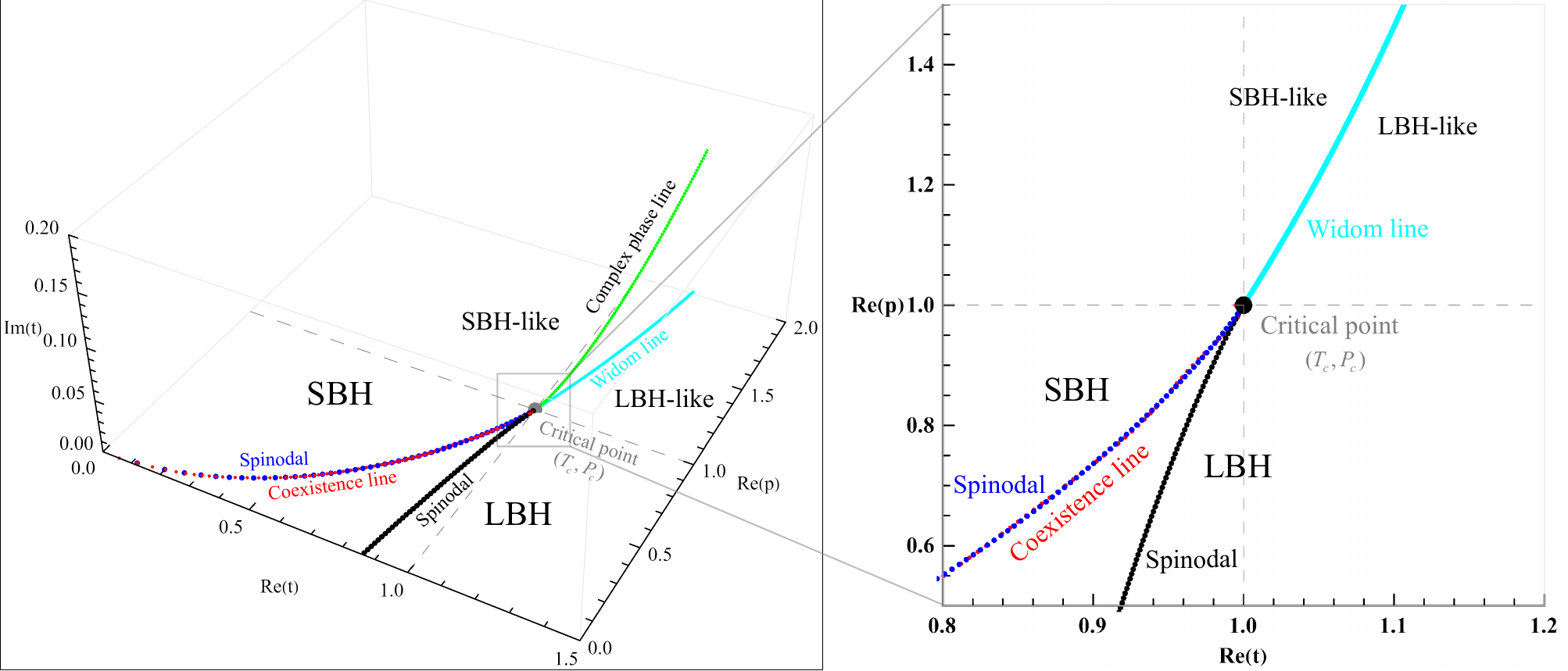}
				\caption{The complex phase diagram of the 5d charged Gauss-Bonnet AdS black hole and corresponding supercritical phenomena. 
				}
				\label{fig3}	
			\end{figure}
			
			Following this formalism, we identify the singularities for our 5d system. The condition for $c_p$ divergence (setting the denominator of Eq.~(\ref{seq1}) to zero) yields a polynomial equation in $z$,
			
			\begin{equation}\label{fenbu1}
				p{{z}^{8}}+\left( \frac{6\alpha }{{{r}_{c}}^{2}}p-\frac{3}{8\pi {{P}_{c}}{{r}_{c}}^{2}} \right){{z}^{6}}+\frac{3\alpha }{4\pi {{P}_{c}}{{r}_{c}}^{4}}{{z}^{4}}+\frac{5{{Q}^{2}}}{8\pi {{P}_{c}}{{r}_{c}}^{6}}{{z}^{2}}+\frac{3\alpha {{Q}^{2}}}{4\pi {{P}_{c}}{{r}_{c}}^{8}}=0.
			\end{equation}

			Although Eq.~(\ref{fenbu1}) is an eighth-order polynomial and thus admits eight roots in the complex $z$-plane, not all roots are physically relevant. Our numerical analysis confirms that four of these eight roots are always located within the non-physical region ( $|z| < z_{\text{min}}(\text{5d})$, corresponding to the gray shaded area in Fig.~\ref{fig2}) for all pressures. These roots are therefore discarded from our thermodynamic analysis.
			
			The remaining four roots satisfy the physical constraints from Eq.~(\ref{prg1}) and the positive temperature condition $z > z_{\text{min}}(\text{5d})$ and are considered the physical roots. As shown in Fig.~\ref{fig2}, the distribution of these four physical roots exhibit two distinct behaviors depending on the pressure.
			
			\begin{itemize} 
				\item For $p \le 1$, the roots lie on the real axis ($z \in \mathbb{R}$). These roots correspond to the phase transition and the metastable phases. As $p \to 1$ from below, the real roots associated with the phase transition merge at the critical point $z_c$. 
				\item For $p > 1$, the roots move off the real axis into the complex plane, forming two complex conjugate pairs. These roots govern the supercritical behavior.
			\end{itemize}

			According to Ref.\cite{Xu:2025jrk}, the complex roots in the first quadrant (those with $\Re(z) > 0$ and $\Im(z) > 0$) determine the complex phase line. Using the roots from Eq.~(\ref{fenbu1}) and the temperature relation from Eq.~(\ref{seq1}), we build the complete complex phase diagram shown in Fig.~\ref{fig3}.  We emphasize that the components involving imaginary axes (e.g., $\Im(p)$ or $\Im(t)$) are a mathematical construction, as the physical significance lies entirely in the projection of these complex singularities onto the real $(\Re(p), \Re(t))$ plane.  This projection, shown in the Fig.~\ref{fig3} right panel, defines the Widom line.
			
			The right panel provides a magnified view near the critical point. The main features of this 5d phase diagram are as follows
			
			\begin{itemize} 
				\item Coexistence line (red) indicates the first-order phase transition between the SBH and LBH phases, terminating at the critical point $(T_c, P_c)$. 
				\item Spinodal lines (black and blue) corresponding to the real singularities for $p < 1$, these lines meet at the critical point and define the limits of metastable states. 
				\item Complex phase line (green) is determined by the complex singularities for $p > 1$ (those in the first quadrant of Fig.~\ref{fig2}) and extends from the critical point into the complex space. 
				\item Widom line (cyan) is the projection of the complex phase line onto the real $(\Re(p), \Re(t))$ plane. It emanates from the critical point and separates the supercritical region into a SBH-like phase and a LBH-like phase. 
			\end{itemize}
			
			The phase pattern for the 5d case is similar to that of the 4d charged AdS black hole\cite{Xu:2025jrk}, featuring a single Widom line. This similarity suggests that exploring the 6d case next could reveal new supercritical behaviors, especially around the possible triple point.
			
			\section{Phase structure and supercriticality in six-dimensions}\label{IV}
			Following this, we further analyze the 6d case, which exhibits a richer and more complex thermodynamic structure.
			\subsection{Reduced equations of state and critical behavior}
			
			Using the reduction scheme defined in (\ref{6dr}) with $Q=1, \alpha=3.05$, and reduced relative to the 6d triple point, the triple point parameters are\cite{Wei_2022}
			\begin{equation}
				P_t \approx 0.00639, \quad T_t \approx 0.0642, \quad r_t \approx 1.888, \quad G_t \approx 3.271.
			\end{equation}
			The 6d system also exhibits three critical points, with the corresponding pressure $P$ and horizon radius $r_h$ given as follows
			\begin{equation}
				\begin{aligned}
					\text{Critical point 1: }\quad & P_{c}^{(1)} \approx 0.00663, \quad {r}_{c}^{(1)} \approx 1.452, \\
					\text{Critical point 2: }\quad & P_{c}^{(2)} \approx  0.00628, \quad {r}_{c}^{(2)} \approx 1.929, \\
					\text{Critical point 3: }\quad & P_{c}^{(3)} \approx  0.00647, \quad{r}_{c}^{(3)} \approx 2.724.
				\end{aligned}
			\end{equation}

			The reduced thermodynamic quantities then become
			\begin{equation}\label{seq2}
				\begin{split}
					&t=\frac{8\pi p{{P}_{t}}{{r}_{t}}^{8}{{z}^{8}}+6{{r}_{t}}^{6}{{z}^{6}}+2\alpha {{r}_{t}}^{4}{{z}^{4}}-{{Q}^{2}}}{8\pi {{r}_{t}}^{7}{{T}_{t}}{{z}^{7}}+16\pi \alpha {{r}_{t}}^{5}{{T}_{t}}{{z}^{5}}},  \\ 	
					&g=\frac{1}{180{{G}_{t}}{{r}_{t}}^{3}{{z}^{3}}\left( {{r}_{t}}^{2}{{z}^{2}}+2\alpha \right)} \left[ 5\pi {{Q}^{2}}\left( 7{{r}_{t}}^{2}{{z}^{2}}+20\alpha \right) - 6\pi {{r}_{t}}^{4}{{z}^{4}}\left( 4p{{P}_{t}}\pi {{r}_{t}}^{6}{{z}^{6}}+5{{r}_{t}}^{2}{{z}^{2}}\alpha \right) \right. \\
					&\left. \quad - 6\pi {{r}_{t}}^{4}{{z}^{4}} \left( -20{{\alpha }^{2}}+{{r}_{t}}^{4}{{z}^{4}}\left( -5+48p{{P}_{t}}\pi \alpha \right) \right) \right], \\
					&{{c}_{p}}=\frac{8{{P}_{t}}{{\pi }^{2}}{{r}_{t}}^{2}{{z}^{2}}{{\left( {{r}_{t}}^{2}{{z}^{2}}+2\alpha  \right)}^{2}}\left( -{{Q}^{2}}+6{{r}_{t}}^{6}{{z}^{6}}+8p{{P}_{t}}\pi {{r}_{t}}^{8}{{z}^{8}}+2{{r}_{t}}^{4}{{z}^{4}}\alpha  \right)}{3 D}, \\ 
				\end{split}		
			\end{equation}
			where
			\begin{equation}
				D=\left( 8p{{P}_{t}}\pi {{r}_{t}}^{10}{{z}^{10}}+6{{r}_{t}}^{6}{{z}^{6}}\alpha -4{{r}_{t}}^{4}{{z}^{4}}{{\alpha }^{2}}+{{Q}^{2}}\left( 7{{r}_{t}}^{2}{{z}^{2}}+10\alpha  \right)+6{{r}_{t}}^{8}{{z}^{8}}\left( -1+8p{{P}_{t}}\pi \alpha  \right) \right).
			\end{equation}

			The thermodynamic behavior in 6d is summarized in Fig.\ref{fig4}. As in the 6d case, the temperature $t$ also turns negative at small $z$ (Fig.~\ref{fig4a}). To ensure physical consistency, we again restrict to states with $t > 0$, which imposes the bounds
			\begin{equation}\label{rg2}
				0 < p \le \frac{5}{16\pi\alpha} / P_{t}, \quad z > z_{\text{min}}(\text{6d}).
			\end{equation}
			
			The Gibbs free energy shows a complex structure that indicates multiple phase transitions (Fig.~\ref{fig4b}). At pressures below  the triple point ($p < 1$), the $g - t$ curve evolves from a single swallowtail into a double swallowtail as pressure increases. This signals the emergence of a stable IBH phase. At the triple point ($p = 1$), three stable branches meet, confirming the coexistence of SBH, IBH, and LBH. This is in basically consistent with the detailed phase diagrams established in Refs.\cite{Wei:2014hba,Frassino:2014pha,Wei_2022}.

			\begin{figure}[htbp]
				\centering
				\begin{subfigure}[b]{0.48\textwidth}
					\includegraphics[width=\textwidth]{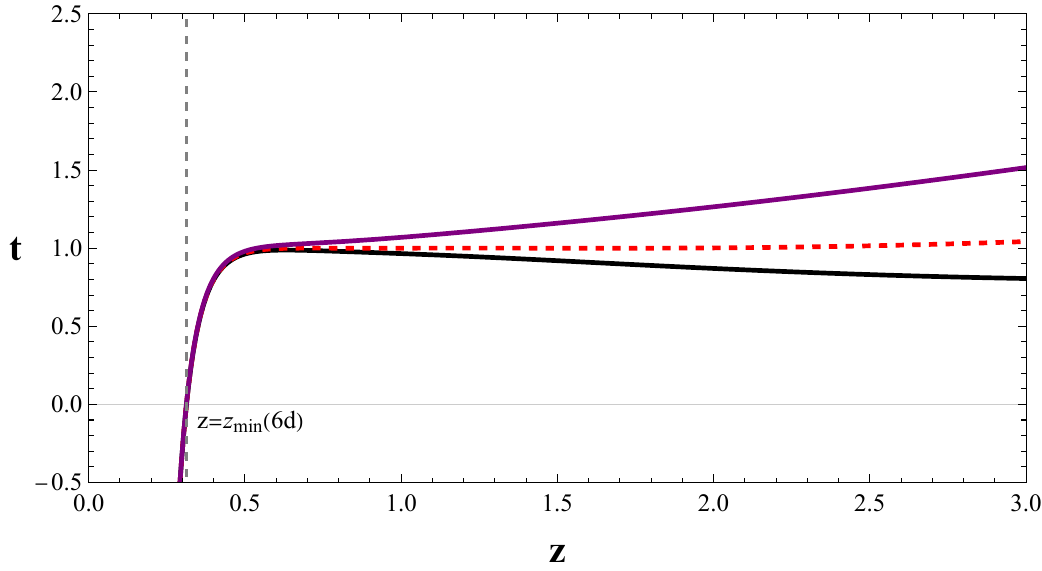}
					\caption{Temperature ($t$) vs. Horizon radius ($z$)}
					\label{fig4a}
				\end{subfigure}
				\hfill
				\begin{subfigure}[b]{0.48\textwidth}
					\includegraphics[width=\textwidth]{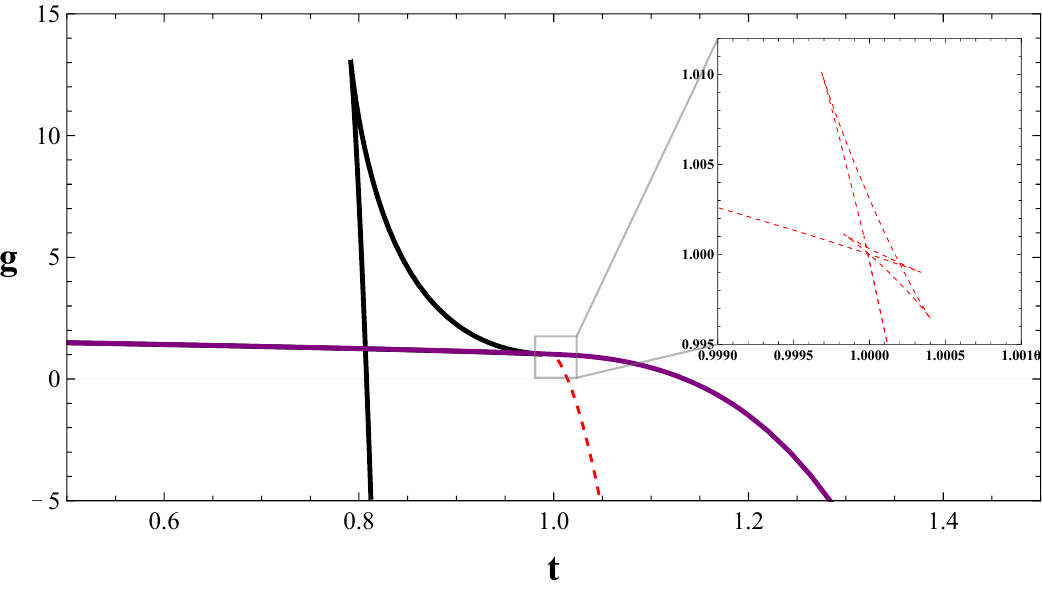}
					\caption{Gibbs free energy ($g$) vs. Temperature ($t$)}
					\label{fig4b}
				\end{subfigure} \\
				
				\begin{subfigure}[b]{0.48\textwidth}
					\includegraphics[width=\textwidth]{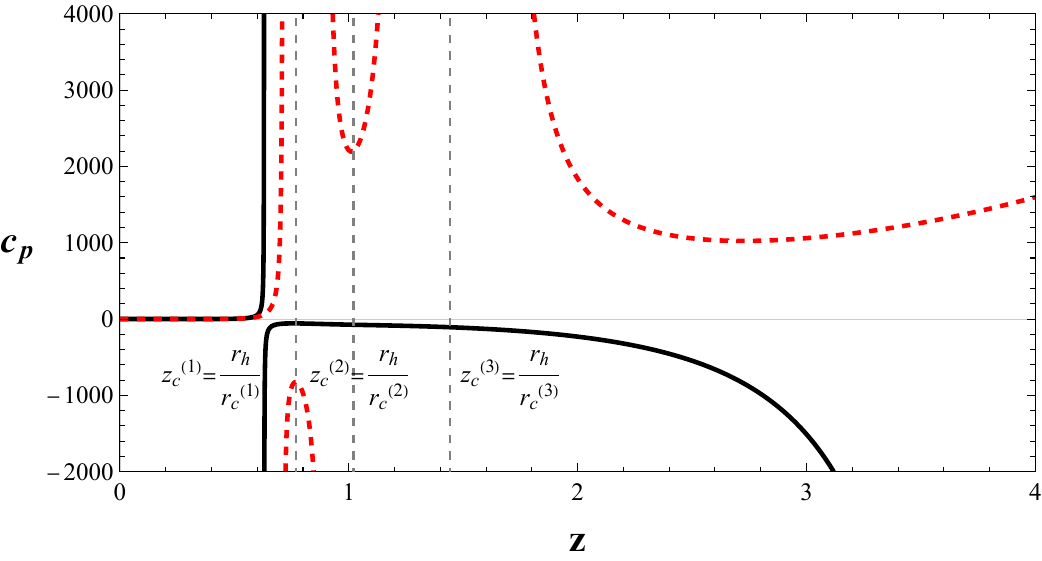}
					\caption{Heat capacity ($c_p$) for $p \le 1$}
					\label{fig4c}
				\end{subfigure}
				\hfill
				\begin{subfigure}[b]{0.48\textwidth}
					\includegraphics[width=\textwidth]{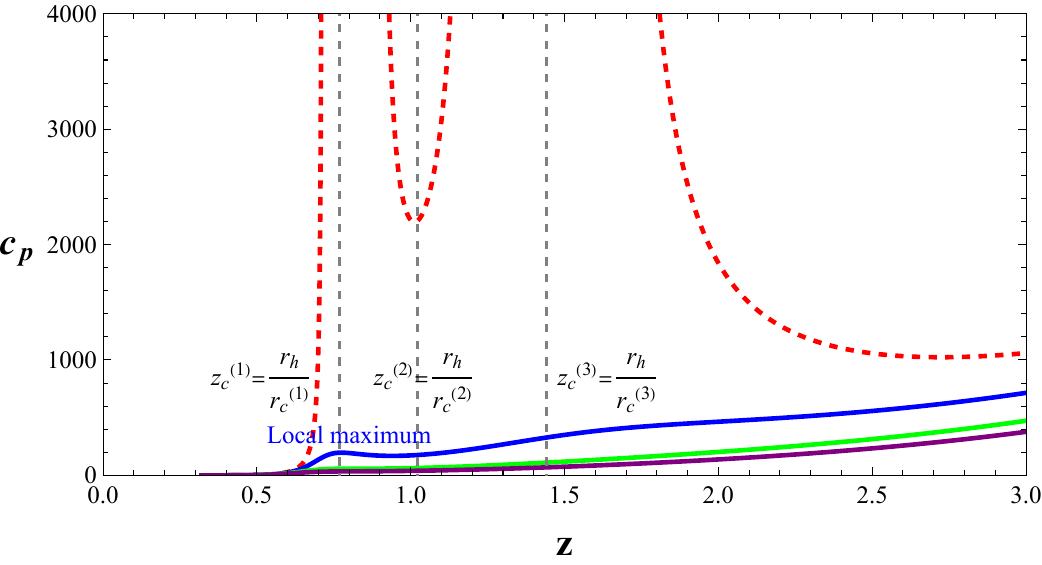}
					\caption{Heat capacity ($c_p$) for $p \ge 1$}
					\label{fig4d}
				\end{subfigure}
				
				\caption{Thermodynamic behavior in 6d charged Gauss-Bonnet AdS black hole ($Q=1, \alpha=3.05$). Curves correspond to different reduced pressures: $p=0.50$ (black), $p=1.00$ (red dashed, triple point), $p=1.20$ (blue), $p=1.60$ (green), $p=2.00$ (purple). The vertical dashed line in (a) indicates the physical boundary $z_{\text{min}}(\text{6d}) \approx 0.313$, below which $t<0$. The vertical dashed lines in (c) at $z_c^{(1)} \approx 0.769$, $z_c^{(2)} \approx 1.027$, and $z_c^{(3)} \approx 1.443$ mark the radii of the three critical points, respectively.}
				\label{fig4}
				
			\end{figure}

			This complex phase behavior is further reflected in the heat capacity $c_p$ (Fig.~\ref{fig4c}). At pressures below the respective critical points (e.g., $p=0.50$, black curve), $c_p$ exhibits multiple divergences, indicating several distinct spinodal lines. As the pressure approaches the critical values, these divergence points merge, culminating at the critical radius $z_c^{(1)}$, $z_c^{(2)}$, and $z_c^{(3)}$, which are marked by the vertical dashed lines in Fig.~\ref{fig4c}. For pressures above the critical regime (Fig.~\ref{fig4d}), these divergences are replaced by finite local maxima, signaling the crossover into the supercritical region. As established in Section~\ref{III}, these local maxima are insufficient for rigorously defining the Widom lines, necessitating the Lee-Yang approach.

			\subsection{Complex phase diagram and multiple Widom lines}
			
			The singularity condition for the Gibbs free energy in the 6d case, determined by the divergence of the heat capacity $c_p$ (\ref{seq2}), which yields a tenth-order polynomial equation.
			
			\begin{equation}\label{fenbu2}
				p{{z}^{10}}+\left( \frac{6\alpha }{{{r}_{t}}^{2}}p-\frac{3}{4\pi {{P}_{t}}{{r}_{t}}^{2}} \right){{z}^{8}}+\frac{3\alpha }{4\pi {{P}_{t}}{{r}_{t}}^{4}}{{z}^{6}}-\frac{{{\alpha }^{2}}}{2\pi {{P}_{t}}{{r}_{t}}^{6}}{{z}^{4}}+\frac{7{{Q}^{2}}}{8\pi {{P}_{t}}{{r}_{t}}^{8}}{{z}^{2}}+\frac{5\alpha {{Q}^{2}}}{4\pi {{P}_{t}}{{r}_{t}}^{10}}=0.
			\end{equation}
			
			The singularity condition Eq.~(\ref{fenbu2}) is a tenth-order polynomial. Applying the physical constraints from Eq.~(\ref{rg2}) leave eight physical roots, as this process excludes two solutions within the non-physical gray region ($|z| < z_{\text{min}}(\text{6d})$). The distribution of these physical roots is shown in Fig.~\ref{fig5}.

			The behavior of these physical roots is significantly more complex than in the 5d case. As shown in Fig.~\ref{fig5}, the system exhibits a staggered transition from real to complex roots, governed by the three critical points. These  roots merge  at three distinct locations on the real axis: $\Re(z) = z_c^{(1)}, z_c^{(2)}, z_c^{(3)}$. These correspond to the reduced critical pressures $p_c^{(1)} \approx 1.038$, $p_c^{(2)} \approx 0.983$, and $p_c^{(3)} \approx 1.013$, respectively.
			
		    According to the theory, the complex roots in the first quadrant determine the complex phase lines. These complex and real roots are then used to construct the full phase diagram shown in Fig.~\ref{fig6}, which is qualitatively different from the 5d case. The main features are as follows

			\begin{itemize}
				\item Coexistence lines (yellow, orange, red) show the boundaries where first-order phase transitions occur between SBH/LBH, SBH/IBH, and IBH/LBH, respectively. When these lines are crossed, thermodynamic quantities undergo discontinuous changes. 
				\item Spinodal lines (black, blue, pink, magenta) correspond to the four segments of Gibbs free energy singularities on the positive real $z$-axis for $p<1$. These lines pass through three critical points ( $( T^{(1)}_c, P^{(1)}_c),( T^{(2)}_c, P^{(2)}_c),( T^{(3)}_c, P^{(3)}_c)$ ),  indicating the boundaries of each metastable phase.
				\item Complex phase lines (green, brown) arise from the analytic continuation of Gibbs free energy singularities into the first quadrant of the complex $z$-plane for $p>1$. They reflect phase transition behavior in the supercritical regime.
				\item Widom lines (cyan, purple) are defined as the real-plane projections of the complex phase lines. These smooth crossover boundaries separate the supercritical region into three domains (SBH-like, IBH-like, and LBH-like) across which thermodynamics evolves continuously, without singularities.
			\end{itemize}

			\begin{figure}[htbp]
				\centering
				\includegraphics[width=110 mm]{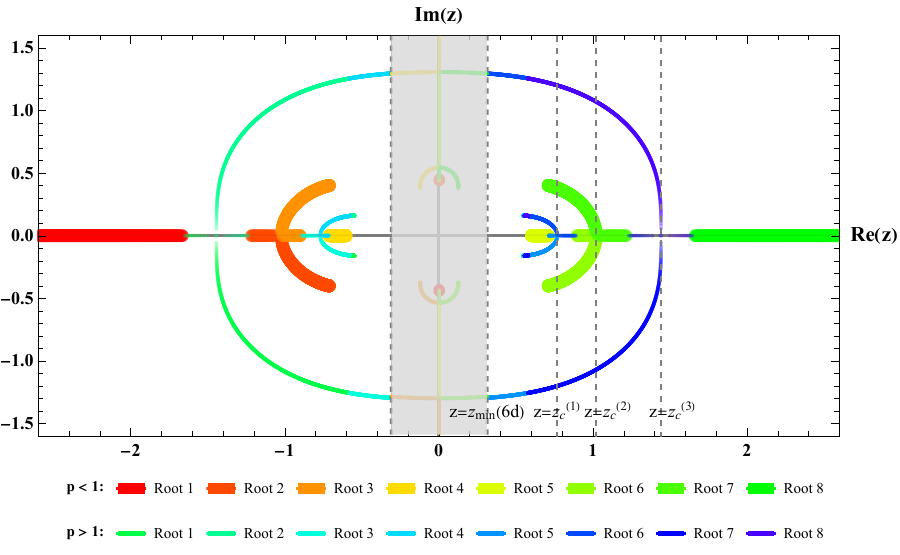}
				\caption{The singularity distribution for the 6d charged Gauss-Bonnet AdS black hole. The distinct colors denote the four physical roots of Eq.~(\ref{fenbu2}). The gray shaded region ($|z| < z_{\text{min}}(\text{6d})$) is non-physical, and roots within it are excluded. The vertical dashed line marks the critical point $z_c^{(1)}, z_c^{(2)}, z_c^{(3)}$ on the real axis. 
				}
				\label{fig5}	
			\end{figure}

			\begin{figure}[htbp]
				\centering
				\includegraphics[width=150mm]{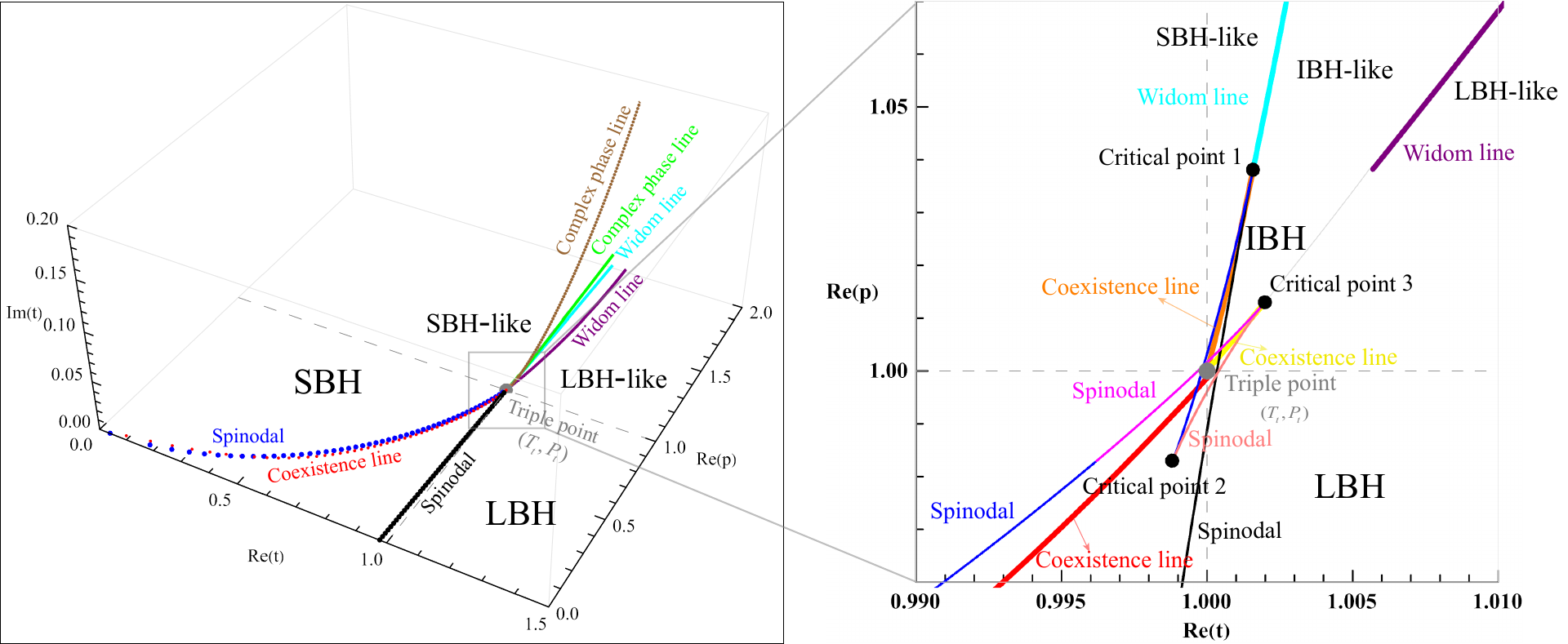}
				\caption{Complex phase diagram of the 6d charged Gauss–Bonnet AdS black hole. Two Widom lines partition the supercritical region into three distinct phase-like sectors.}
				\label{fig6}
			\end{figure}
			
			These findings demonstrate that the triple point, a distinctive feature of the 6d Gauss–Bonnet case, it leaves a direct imprint on the supercritical state. In contrast, lower-dimensional systems without a triple point do not support such richly organized supercritical phenomenology. As a result, we have refined the phase diagram of Gauss–Bonnet AdS black holes.

			\section{Discussion and conclusion}\label{V}
			In this work, we utilize the Lee-Yang phase transition theory to extend the thermodynamic analysis of charged Gauss-Bonnet anti-de Sitter black holes into the supercritical regime. By analytically extending thermodynamic quantities to the complex domain and studying the singularities of the Gibbs free energy (these singularities correspond to the Lee-Yang zeros of the partition function), we construct a complete complex phase diagram, revealing the unique thermodynamic structure of the black hole in the supercritical region.
			
			Based on the differences in phase structures caused by the Gauss-Bonnet coupling across different dimensions, the study demonstrates that the thermodynamic behavior in the supercritical region of five and six-dimensional charged Gauss-Bonnet AdS black holes changes dramatically. The five-dimensional system exhibits a first-order phase transition between small and large black holes, characterized by a single critical point. Its supercritical area is separated into two subregions, the small black hole-like phase and the large black hole-like phase, by a single Widom line that serves as a dynamic boundary. This behavior is qualitatively compatible with that of four-dimensional charged AdS black holes, demonstrating that VdW-like systems exhibit a common supercritical pattern when analyzed via the Lee-Yang theory.

			In contrast, the six-dimensional system shows more complex thermodynamic behavior.  For a given Gauss–Bonnet coupling, a triple point appears where small, intermediate, and large black hole phases coexist in equilibrium.  The projections of the Gibbs free energy onto the real phase plane create two separate Widom lines in the supercritical region, where this multiphase structure remains. These lines together form three regions in the supercritical region, representing small-, intermediate-, and large-black hole-like phases.  This suggests that the distribution of phases becomes more complicated due to the ongoing critical and multicritical processes in the supercritical state.
			
			Our analysis focuses on $\alpha=3.05$, a value permitting a triple point.  A natural question concerns the generality of the resulting two Widom lines structure.  We posit that this structure is a necessary consequence of the underlying phase behavior.  The existence of three coexisting phases (SBH/IBH/LBH) at the triple point inherently partitions the supercritical domain into three corresponding sectors (SBH-like, IBH-like, and LBH-like).  Consequently, two distinct Widom lines are required to delineate these three sectors. This implies that the existence of two Widom lines, rather than their specific paths, should be a general feature for any $\alpha$ value within the parameter range that supports a triple point. 
			
			Conversely, should $\alpha$ be varied outside this range, the system reverts to a single critical point.  We then conjecture the two Widom lines would merge, causing the supercritical structure to degenerate back to the single line case observed in the five-dimensional system. A comprehensive scan of the $\alpha$ parameter space to explicitly map this structural evolution warrants detailed investigation in future work.
			
			To sum up, our results substantially extend the phase diagram of Gauss–Bonnet AdS black holes. We show that their supercritical region is not a featureless domain but is instead partitioned into distinct sectors by Widom lines. These lines, rooted in complex singularities of the Gibbs free energy, arise from the fundamental analytical properties of the partition function. This provides further evidence for the validity of the Lee–Yang approach in black hole thermodynamics and highlights the crucial roles of dimensions and higher-curvature corrections. Future research could explore black hole systems in more modified gravity theories within the supercritical regime and investigate the characteristics of these phases

			\section*{Acknowledgement}
			This work is supported by the National Natural Science Foundation of China (Grant Nos.12275216, 12105222, 12247103).
		\end{spacing}

\end{document}